\documentclass{article}

\usepackage{amsmath}
\usepackage{latexsym}
\usepackage{amssymb}
\usepackage{bm}
\usepackage{graphics,epstopdf}
\usepackage{color}
\usepackage{hyperref}

\usepackage{newlfont}
\usepackage{amsfonts}
\usepackage{amsthm}
\usepackage{graphicx}
\usepackage{epsfig}

\newcommand{\ket}[1]{|{#1}\rangle}
\newcommand{\bra}[1]{\langle{#1}|}
\newcommand{\braket}[2]{\langle {#1} | {#2} \rangle}
\newcommand{\ketbra}[2]{|{#1} \rangle \langle {#2} |}
\renewcommand{\t}[1]{\textrm{#1}}
\newcommand{\com}[1]{[\textsc{#1}]}
\usepackage{times}

\newcommand*{\Perm}[2]{{}^{#1}\!P_{#2}}%
\newcommand*{\Comb}[2]{{}^{#1}C_{#2}}%
\usepackage[up]{subfigure}

\newcommand{\be}{\begin{equation}}
\newcommand{\ee}{\end{equation}}
\newcommand{\bc}{\begin{center}}
\newcommand{\ec}{\end{center}}
\newcommand{\bea}{\begin{eqnarray}}
\newcommand{\eea}{\end{eqnarray}}
\newcommand{\ba}{\begin{array}}
\newcommand{\ea}{\end{array}}
\newcommand{\tr}{\textcolor{red}}
\newcommand{\tb}{\textcolor{blue}}

\begin{document}

\title{\bf{Measuring Electromagnetic Vector Potential via Weak Value}}

\author{Arun Kumar Pati\\
\small{Quantum Information and Computation Group,}\\
\small{Harish-Chandra Research Institute,}\\
\small{Allahabad 211 019, India}}

\maketitle

\begin{abstract}
Electromagnetic vector potential has physical significance in quantum mechanics as revealed by the Aharonov-Bohm effect for charged 
particles. However, till date it is thought that we cannot measure the vector potential directly as this is not a gauge invariant quantity. 
Contrary to this belief, here we show that one 
can indeed measure the electromagnetic vector potential using the notion of weak value. We show that it is simply the difference between the weak value of the canonical 
momentum of a charged particle in the presence and absence of 
magnetic field. This suggests that the vector potential is not only a physical entity but can be measured directly in experiment.
\end{abstract}



\vskip 1cm


In classical physics, electromagnetic fields have observable consequence on charged particles as manifested through the Lorentz force. The vector and scalar 
potentials from which the electromagnetic fields can be derived have no observable consequence.
However, this is not so in quantum mechanics. With the discovery of the Aharonov-Bohm effect \cite{ab}, the situation has changed. It is possible to 
affect the wave function of a charged particle due to a non zero vector potential even when 
the magnetic field is zero. The non-zero vector potential has ability to modify the phase of the wave function of a charged particle and gives rise to
a shift in the interference pattern \cite{pesh}. Even though classical description of charged particle can be given equivalently using the fields or 
vector and scalar potentials, in quantum mechanics
it is the vector and scalar potentials that enter the Hamiltonian and hence the Schr{\"o}dinger equation. In quantum electrodynamics, the vector and scalar 
potentials are the fundamental quantities instead of the electric and magnetic fields.

   This raises the question whether the vector potential is a physical entity. If it is, can one measure it directly. To the best of our knowledge 
we have not seen any discussion on how to measure the vector potential in quantum theory. The purpose of this note is to show that 
using the concept of weak measurement one can measure the vector potential as seen by a charged particle. We show that the vector potential is nothing but the difference 
between the weak value of the canonical momentum operator in certain pre- and post-selected states in the presence and absence of magnetic field.
Furthermore, we interpret the vector potential as the difference between the best estimate of the canonical momentum of a charged particle 
in the presence and absence of vector potential compatible with the position measurement. Even though our result may look simple, it can change 
the long held view that the vector potential cannot be measured.

The concept of weak value has played a fundamental role in quantum theory.
This was first introduced by Aharonov-Albert-Vaidman
\cite{aha} while investigating the properties of a quantum system in  pre and post-selected ensembles.
 If the system is weakly coupled to an apparatus, then upon post-selection of the system onto a final state, the apparatus variable is shifted by 
 the weak value. Interestingly, the weak value of an observable can have strange properties \cite{duck}.
In general, the weak value can be a complex number and can be large. Moreover, it can take values outside the spectrum of the observable being measured. 
The concept of weak value has found numerous practical applications  in recent years \cite{nw,gj,oh,dixon,hof}. 
This provides us a new tool to look at complementary aspects of quantum quantum world \cite{av,sandu1,sandu2,wu}.
Moreover, using the concept of weak value, one can measure the wave function of a quantum system which was earlier thought to be only a mathematical 
entity \cite{jsl}. Recently, it  has been shown that average of any non-Hermitian operator in any pure quantum state can be measured via the notion of 
weak value \cite{akp}.

 Let us consider a charged particle such as an electron in the presence of vector potential. The Schr{\"o}dinger equation for the electron is given by
 \begin{align}
 i\hbar \frac{d\ket{\Psi}}{dt} = H(\mathbf{p}- \frac{e}{c}\mathbf {A}, \mathbf{x}) \ket{\Psi},
 \end{align}
where $H$ is the Hamiltonian, $e$ is the charge of the electron and $c$ is the speed of light.
Now,  if $\ket{\Psi_0}$ is solution to the Schr{\"o}dinger equaltion in the absence of the vector potential, then the solution to the above equation is given by
\begin{align}
\ket{\Psi} = e^{ \frac{ie}{\hbar c} \int \mathbf{A}(\mathbf{x}'). d\mathbf{x}'} \ket{\Psi_0}
\end{align}

Now we perform the weak measurement of the canonical momentum $\mathbf{p}$ for the charged particle in preselected state $\ket{\Psi}$ and post-selected state $\ket{\mathbf{x}}$.
The weak value of the canonical momentum for the above pre- and post-selection is given by
\begin{align}
\langle \mathbf{p} \rangle_w = \frac{\braket{\mathbf{x}  }{ \mathbf{p} |\Psi }}{\braket{\mathbf{x} }{\Psi} }.
\end{align}
Using (2) this can be expressed as 
\begin{align}
\langle \mathbf{p} \rangle_w = \frac{\braket{\mathbf{x}  }{ \mathbf{p} |\Psi }}{\braket{\mathbf{x}  }{\Psi} } =  
\frac{\braket{\mathbf{x}  }{ \mathbf{p} |\Psi_0 }}{\braket{\mathbf{x}  }{\Psi_0} } + \frac{e}{c} \mathbf{A}
\end{align}
Therefore, we have 
\begin{align}
\langle \mathbf{p} \rangle_w - \langle \mathbf{p} \rangle_w^{(0)}=   \frac{e}{c} \mathbf{A},
\end{align}
where $\langle \mathbf{p} \rangle_w^{(0)} = \frac{\braket{\mathbf{x}  }{ \mathbf{p} |\Psi_0 }}{\braket{\mathbf{x}  }{\Psi_0} } $ is the weak value of 
the canonical momentum in the pre-selected state $\ket{\Psi_0}$ and post-selected state $\ket{\mathbf{x}}$. 
This shows that the vector potential is the difference between the weak value of the canonical 
momentum of a charged particle in the presence and absence of 
magnetic field. This indeed suggests that the vector potential is not only a physical entity but can be measured directly in experiment.

{
In the weak measurement experiment, we take the pre-selected state of an electron $\ket{\Psi_i} = \ket{\Psi_0}$ and an apparatus in the state $\ket{\Phi}$.
The weak measurement can be realized using the interaction between the system and the
measurement apparatus which is governed by the interaction Hamiltonian
\begin{align}
 H_{int}= f(t) \mathbf{p} \otimes M,
\end{align}
where $f(t)$ is the strength of the interaction with $\int f(t)dt= g$, $\mathbf{p}$ is canonical momentum of the charged particle and $M$ is that of the apparatus 
(often called meter variable).
This is the von Neumann model of measurement when the coupling strength is arbitrary. However, if $g$ is small, then we can realize the weak
measurement of the canonical momentum. The interaction Hamiltonian allows the initial state of the system and apparatus $\ket{\Psi_0} \otimes \ket{\Phi}$ to evolve as
\begin{align}
 \ket{\Psi_0} \otimes \ket{\Phi}  \rightarrow e^{-i g \mathbf{p} \otimes M} \ket{\Psi_0} \otimes \ket{\Phi}.
\end{align}
After the weak interaction, we post-select the system in the state $\ket{\mathbf{x} }$ (by performing strong position measurement) with the
postselection
probability given by $p= |\braket{\mathbf{x} }{\psi}|^2 (1 + 2 g Im \langle \mathbf{p} \rangle_w \langle M \rangle)$, where
$\langle M \rangle = \braket{\Phi}{M |\Phi }$ and the weak value of $\mathbf{p} $ is
given by
\begin{align}
\langle \mathbf{p} \rangle_w^{(0)} = \frac{\braket{ \mathbf{x} }{ \mathbf{p} |\Psi_0 }}{\braket{\mathbf{x}  }{\Psi_0} }.
\end{align}
Next, we carry out similar weak measurement for the electron in the presence of magnetic field with the pre-selected state $\ket{\Psi}$ and 
post-selected state $\ket{\mathbf{x}}$. The weak value then will be given by $\langle \mathbf{p} \rangle_w$. From these two weak values one can 
infer the vector potential. By looking at the shift in the pointer state one can measure the real and imaginary parts of these weak values and hence 
the vector potential. Thus the weak measurement scheme provides a means to build a device-- a `potential meter' that can measure the gauge potential.

 In an interesting paper, Hall \cite{hall} has shown that canonical momentum operator for any particle can be decomposed as 
 \begin{align}
 \mathbf{p} = \mathbf{p}_c + \mathbf{p}_{nc},
 \end{align}
 where $\mathbf{p}_c$ is a commuting component with the position observable and $\mathbf{p}_{nc}$ is non-classical part
 that actually does not commute with the position. The commuting component 
$\mathbf{p}_c$ has a spectral decomposition  
\begin{align}
\mathbf{p}_c  = \int d\mathbf{x} \mathbf{p}_c(\mathbf{x}) \ket{\mathbf{x}} \bra{\mathbf{x}}
\end{align}
and $\mathbf{p}_c(\mathbf{x})$ is given 
\begin{align}
\mathbf{p}_c(\mathbf{x}) = \mathrm{Re} \frac{\braket{\mathbf{x} }{ \mathbf{p} |\Psi }}{\braket{\mathbf{x} }{\Psi} }.
\end{align}
Thus, the commuting component of the momentum is nothing but a Hermitian operator with eigenvalues as the real part of the weak value of the 
canonical momentum 
operator in the pre-selected state $\ket{\Psi}$ and post-selected state $\ket{\mathbf{x}}$. Also, it has been shown that $\mathbf{p}_c$ corresponds to the best
possible estimate of the momentum operator, for a given quantum state $\ket{\Psi}$, which is compatible with the position measurement.
Further, it has been proved that the Bayes estimator of canonical 
momentum between pre- and post-selections is equal to the real part of the weak value \cite{lars}. Following this result, we can interpret the vector 
potential as the the difference between the best estimate of the canonical momentum of a charged particle  in the presence and absence of magnetic field 
compatible with the position measurement for the electron.

One may ask the question: since the electromagnetic vector potential is not a gauge invariant quantity, how can one measure this? Before answering this, let us 
note that the wave function of a quantum particle is not a gauge invariant object. Under local gauge transformation we have
$\Psi(\mathbf{x}, t) \rightarrow exp(i\frac{e}{\hbar c} \Lambda(\mathbf{x}, t) )\Psi(\mathbf{x}, t)$.  But still one can measure the wave function with the help of 
the weak measurement in a fixed gauge. Indeed it has been shown that the wave function is nothing but the weak value of the position projection operator between pre- and post-selection 
and this has been measured experimentally \cite{jsl}. Perhaps, one has to distinguish between the meaning of the statement that something has `observable consequence' and something 
as `measurable'. Our result shows that even though the vector potential is not a gauge invariant object, it can be measured via the weak value.
We conjecture that other abelian and non-abelian gauge potentials \cite{yang} may be measured via the weak value for suitable pre- and post-selected states.
This will change the status of gauge potentials from mere mathematical objects to something real and measurable.

\end{document}